\documentclass[conference]{IEEEtran}
\usepackage{amsmath, amssymb, amsfonts}
\usepackage{bm}
\usepackage{graphicx}
\usepackage{cite}
\usepackage{xcolor}
\usepackage{tikz}
\usetikzlibrary{arrows.meta, positioning, shapes.geometric, fit, backgrounds}
\usepackage[acronym,shortcuts]{glossaries}

\title{Learning to Compute on Dirty Paper}

\author{
    \IEEEauthorblockN{Shreesal Shrestha\IEEEauthorrefmark{1}, Kuranage Roche Rayan Ranasinghe\IEEEauthorrefmark{1}, Giuseppe Thadeu Freitas de Abreu\IEEEauthorrefmark{1} and Elza Erkip\IEEEauthorrefmark{2}}
    \IEEEauthorblockA{\IEEEauthorrefmark{1}\textit{School of Computer Science and Engineering, Constructor University, 28759 Bremen, Germany}}
    \IEEEauthorblockA{\IEEEauthorrefmark{2}\textit{Department of Electrical and Computer Engineering, New York University, New York, NY, USA} \\
    Emails: [shrshrestha, kranasinghe, gabreu]@constructor.university, elza@nyu.edu}
    \vspace{-2em}
}

\newacronym{6G}{6G}{sixth generation}
\newacronym{AirComp}{AirComp}{over-the-air computation}
\newacronym{AWGN}{AWGN}{additive white Gaussian noise}
\newacronym{BS}{BS}{base station}
\newacronym{CE}{CE}{cross-entropy}
\newacronym{CSI}{CSI}{channel state information}
\newacronym{DPC}{DPC}{dirty paper coding}
\newacronym{GeLU}{GeLU}{Gaussian error linear unit}
\newacronym{ICC}{ICC}{integrated communication and computing}
\newacronym{i.i.d.}{i.i.d.}{independent and identically distributed}
\newacronym{LMMSE}{LMMSE}{linear minimum mean squared error}
\newacronym{ML}{ML}{maximum likelihood}
\newacronym{MLP}{MLP}{multilayer perceptron}
\newacronym{MSE}{MSE}{mean squared error}
\newacronym{QAM}{QAM}{quadrature amplitude modulation}
\newacronym{QPSK}{QPSK}{quadrature phase shift keying}
\newacronym{SER}{SER}{symbol error rate}
\newacronym{SIMO}{SIMO}{single-input multiple-output}
\newacronym{SNR}{SNR}{signal-to-noise ratio}
\newacronym{SotA}{SotA}{state-of-the-art}
\newacronym{THP}{THP}{Tomlinson-Harashima precoding}
\newacronym{TX}{TX}{transmitter}
\newacronym{RX}{RX}{receiver}

\begin{document}

\maketitle

\begin{abstract}
\Ac{ICC} allows users to transmit data symbols alongside computing symbols for \ac{AirComp} over a shared multiple-access channel, but causes mutual interference between both tasks.
As the computing symbols are known at the users, lattice-based \ac{DPC} can pre-cancel them, as in Lattice \ac{ICC}, leaving the data detection largely free of interference, yet also obscuring them at the receiver, which results in an exponentially complex estimator and an error floor in the computing task.
To overcome this, we propose Learned \ac{ICC}, which replaces the lattice-based precoder with a neural encoder shared by all users, and the receiver with a dual-headed neural decoder that jointly detects the data and directly estimates the target function, both trained end-to-end.
Simulation results show that Learned \ac{ICC} removes the error floor and substantially reduces the computing \ac{MSE} compared to Lattice \ac{ICC}, with a moderate trade-off in data detection.
%
\end{abstract}

\glsresetall

\section{Introduction}
\label{sec:intro}

Beyond connectivity, \ac{6G} networks \cite{Saad_2020, Tataria_2021} are envisioned to support distributed intelligence at the network edge, where massive numbers of devices must not only exchange data, but also jointly process it, as in federated learning and large-scale sensor networks \cite{Li_FedLearning_Chal, Yang_FedLearning_2020, Chen_OTAC_IoT_2018}.
Among its enablers, \ac{AirComp} exploits the superposition property of the multiple-access channel to compute nomographic functions of distributed data directly over the air \cite{Nazer_2007, Liu_2020, Wang_2024, Ando_ITJ2025, Ando_CAMSAP2023, Rou_ICASSP2026}.
Building on this principle, \ac{ICC} performs such computation simultaneously with data transmission \cite{Ranasinghe_2026, Du_2024}.

Sharing the channel, however, causes the two tasks to interfere.
Existing \ac{ICC} receivers \cite{Ranasinghe_2026} treat the computing signal as noise when detecting the data, and estimate the target function from the residual, leaving both tasks interference-limited.
This approach overlooks that the computing symbols are known non-causally at the users, which is precisely the setting of Costa's \ac{DPC} \cite{Costa1983}, showing that, over a Gaussian noise channel, such interference need not reduce the capacity.
Practical realizations of \ac{DPC} range from \ac{THP}, which applies a scalar modulo operation, to its higher-dimensional generalization via nested lattice codes \cite{KusmeTHP_2005, Damico_2008, Zamir_2002, Erez_2005, Kochman_2009}.
The latter was recently applied to \ac{ICC} in a scheme referred to as Lattice \ac{ICC} hereafter \cite{Ranasinghe_DPC_2026}, letting the data detection approach its interference-free performance.
However, cancelling the computing symbols entirely also obscures them at the \ac{RX}, such that their recovery requires a \ac{ML} search of exponential complexity in the number of users, and suffers from an error floor that has so far only been lowered by repeated transmissions over multiple time slots.

In parallel, end-to-end learning has emerged as a powerful tool for jointly designing transmitters and receivers \cite{Ye_2020, Oshea2017, Aoudia_2019}.
For \ac{DPC}, it has been shown to recover the modulo structures of \ac{THP} and lattice-based precoding, while outperforming them, particularly at low \ac{SNR} \cite{Ozyilkan_2025}.
These works, however, treat the known interference as a nuisance, whereas in \ac{ICC}, the \ac{RX} must also recover a function of it.

In this paper, we bring both lines of work together and propose Learned \ac{ICC}, a learning-based approach to \ac{ICC} on dirty paper.
Rather than fixing the precoder by design, a neural encoder learns how to write the data around the computing symbols, that is, to \emph{compute on dirty paper}, optimizing the balance between protecting the data and preserving the computing symbols directly through the training objective.
The encoder is shared by all users and requires neither \ac{CSI} nor the symbols of other users, while at the \ac{RX}, a \ac{LMMSE}-aided dual-headed neural decoder detects the data and estimates the target function directly, avoiding the exponential search of Lattice \ac{ICC}.
Our simulations show that this learned balance removes the error floor in the computing \ac{MSE} at a moderate cost in data detection performance.


\section{System Model}
\label{sec:system_model}

Consider a \ac{SIMO} uplink in which $K$ single-antenna users simultaneously transmit to a \ac{BS} equipped with $N$ antennas.
Transmission is organized in blocks of $T$ time slots, indexed by $t \in \{1, \dots, T\}$.
In every slot, each user $k \in \{1, \dots, K\}$ transmits a data symbol $v_k^{(t)} \in \mathcal{D}$, where $\mathcal{D}$ denotes a discrete data constellation with cardinality $D = |\mathcal{D}|$.
In addition, each user holds a computing symbol $s_k \in \mathcal{S}$, where $\mathcal{S}$ denotes a discrete computing constellation, which contributes to an \ac{AirComp} task at the \ac{BS}.
Both constellations are specified in Section~\ref{sec:codpc}.

Under the assumption of perfect symbol synchronization amongst users, the signal received at the \ac{BS} in slot $t$ is given by
\begin{equation}
    \bm{y}^{(t)} = \sum_{k=1}^{K} \bm{h}_k\, x_k^{(t)} + \bm{w}^{(t)} \in \mathbb{C}^{N \times 1},
    \label{eq:rx_signal}
\end{equation}
where $x_k^{(t)} \in \mathbb{C}$ is the multifunctional transmit signal of user $k$, carrying both its data and computing symbols, $\bm{h}_k \in \mathbb{C}^{N \times 1}$ is the channel vector between user $k$ and the \ac{BS}, whose entries are \ac{i.i.d.} as $\mathcal{CN}(0, 1)$, and $\bm{w}^{(t)} \sim \mathcal{CN}(\mathbf{0}, \sigma_w^2 \mathbf{I}_N)$ is the \ac{AWGN} vector with noise variance $\sigma_w^2$.

Following the block Rayleigh fading model typically assumed in the \ac{AirComp} literature \cite{Nazer_2007, Liu_2020, Ando_CAMSAP2023, Ando_ITJ2025, Wang_2024}, the channel matrix $\bm{H} \triangleq [\bm{h}_1, \dots, \bm{h}_K] \in \mathbb{C}^{N \times K}$ remains constant over the $T$ slots of a block.
Perfect \ac{CSI} is assumed at the \ac{BS}, while the users have no \ac{CSI}.
Each user is subject to an average power constraint $\mathbb{E}\big[|x_k^{(t)}|^2\big] \leq P$, where $P$ denotes the average transmit power per user, and the \ac{SNR} is accordingly defined as $\mathrm{SNR} \triangleq P / \sigma_w^2$.

Within a block, the data symbols $v_k^{(t)}$ are drawn independently in every slot, retaining the full communication rate, while the computing symbol $s_k$ is held fixed across all $T$ slots \cite{Ranasinghe_DPC_2026}.
The \ac{BS} thus collects $T$ observations of the same computing signal, which can be averaged to suppress the noise in the estimate of the target function.

\subsection{AirComp Operation}

The goal of \ac{AirComp} is to evaluate a target function of the computing symbols at the \ac{BS}, which, for the general class of nomographic functions \cite{Wang_2024, Liu_2020}, takes the form
\begin{equation}
    f(\bm{s}) = \phi\Bigg(\sum_{k=1}^{K} \psi_k(s_k)\Bigg),
    \label{eq:nomographic}
\end{equation}
where $\bm{s} \triangleq [s_1, \dots, s_K]^\mathsf{T} \in \mathcal{S}^K$ collects the computing symbols of all users, $\psi_k(\cdot)$ denotes the pre-processing function applied at user $k$, and $\phi(\cdot)$ denotes the post-processing function applied at the \ac{BS}.
While many nomographic functions are considered in the \ac{AirComp} literature \cite{Wang_2024, Liu_2020, Nazer_2007}, we adopt the arithmetic sum for ease of exposition, i.e., both $\psi_k(\cdot)$ and $\phi(\cdot)$ are the identity map, such that
\begin{equation}
    f(\bm{s}) = \sum_{k=1}^{K} s_k.
    \label{eq:sum}
\end{equation}

\subsection{Conventional ICC Transmission}

Under the conventional \ac{ICC} paradigm \cite{Ranasinghe_2026}, each user simply superimposes its data and computing symbols, such that
\begin{equation}
    x_k^{(t)} = v_k^{(t)} + s_k.
    \label{eq:sota_tx}
\end{equation}

Stacking the data and transmit symbols of all users into $\bm{v}^{(t)} \triangleq [v_1^{(t)}, \dots, v_K^{(t)}]^\mathsf{T}$ and $\bm{x}^{(t)} \triangleq [x_1^{(t)}, \dots, x_K^{(t)}]^\mathsf{T}$, respectively, the received signal in \eqref{eq:rx_signal} can be compactly written as
\begin{equation}
    \bm{y}^{(t)} = \bm{H}\bm{x}^{(t)} + \bm{w}^{(t)} = \bm{H}\big(\bm{v}^{(t)} + \bm{s}\big) + \bm{w}^{(t)}.
    \label{eq:sota_rx}
\end{equation}

The \ac{BS} then aims to decode the data symbols $\bm{v}^{(t)}$ in every slot and to estimate $f(\bm{s})$ from the $T$ observations of the block.
The \ac{SotA} receiver \cite{Ranasinghe_2026} first decodes the data while treating the computing signal as noise, and then estimates $f(\bm{s})$ via a combiner.
As the two signals overlap in \eqref{eq:sota_rx}, however, each task is degraded by the other, which limits such receiver-side designs.

This motivates addressing the interference at the \ac{TX} instead, where each user knows its computing symbol $s_k$ non-causally.
From the perspective of the data, $s_k$ thus constitutes interference known at the \ac{TX}, which can be pre-cancelled via \ac{DPC} \cite{Costa1983, Ranasinghe_DPC_2026}, such that the data symbols reach the \ac{BS} largely free of computing interference.

\section{Lattice-Based DPC for ICC}
\label{sec:codpc}

We briefly review Lattice \ac{ICC}, proposed in \cite{Ranasinghe_DPC_2026} under the name Computing on Dirty Paper, which serves as the primary baseline of this work.
\vspace{-1em}

\subsection{Nested Lattices and Constellations}

Lattice \ac{ICC} relies on a pair of nested lattices.
Both lattices are scaled versions of the Gaussian integers $\mathbb{Z}[j] \triangleq \{a + jb \mid a, b \in \mathbb{Z}\}$, which form the integer grid in the complex plane.
The coarse lattice $\Lambda_c \triangleq \Delta\,\mathbb{Z}[j]$ partitions the complex plane into square cells of side $\Delta$, and its Voronoi region $\mathcal{V}(\Lambda_c) \triangleq [-\tfrac{\Delta}{2}, \tfrac{\Delta}{2})^2$, the cell around the origin, defines the region into which the modulo operation folds the transmit signal, thereby bounding the transmit power.
The fine lattice $\Lambda_f \triangleq \delta\,\mathbb{Z}[j]$, with spacing $\delta \triangleq \Delta / \sqrt{D}$, is a denser grid that contains $\Lambda_c$, such that each coarse cell holds exactly $D$ fine lattice points.
The two lattices thus play complementary roles where the fine lattice carries the information, while the coarse lattice bounds the power. 

Accordingly, the data constellation consists of the fine lattice points within the Voronoi region, $\mathcal{D} = \Lambda_f \cap \mathcal{V}(\Lambda_c)$, which corresponds to a $D$-ary \ac{QAM} constellation\footnote{For even $\sqrt{D}$, the fine lattice is shifted by $\tfrac{\delta}{2}(1+j)$, such that the constellation is symmetric around the origin.}.
The computing constellation $\mathcal{S} = e^{j\pi/4}\mathcal{D}$ is a $45^\circ$-rotated version of $\mathcal{D}$, such that data and computing symbols do not overlap.
Their average powers are denoted by $E_d \triangleq \mathbb{E}\big[|v_k^{(t)}|^2\big]$ and $E_s \triangleq \mathbb{E}\big[|s_k|^2\big]$, respectively.
\vspace{-1em}

\subsection{Transmitter Design}

Each user pre-cancels its computing symbol from its data symbol and folds the result back into the Voronoi region to bound the transmit power, yielding the precoded data symbol
\begin{equation}
    \tilde{v}_k^{(t)} = \mathrm{modL}\big(v_k^{(t)} - s_k - u_k^{(t)},\, \Delta\big),
    \label{eq:dpc_precoding}
\end{equation}
where $u_k^{(t)}$ is a public dither, drawn uniformly over $\mathcal{V}(\Lambda_c)$ in every slot and known to both the \ac{TX} and the \ac{RX}, and the complex modulo operation is defined as
\begin{align}
    & \mathrm{modL}(a, b) \triangleq \label{eq:modL} \\
    & \quad \Big(\mathrm{mod}\big(\Re\{a\} + \tfrac{b}{2},\, b\big) - \tfrac{b}{2}\Big) + j\Big(\mathrm{mod}\big(\Im\{a\} + \tfrac{b}{2},\, b\big) - \tfrac{b}{2}\Big), \nonumber
\end{align}
with $\mathrm{mod}(\cdot, \cdot)$ denoting the standard modulo operation.
For $b = \Delta$, $\mathrm{modL}(\cdot, \Delta)$ thus folds its argument into $\mathcal{V}(\Lambda_c)$.
The precoded data symbol is then superimposed onto the computing symbol, yielding the transmit signal
\begin{equation}
    x_k^{(t)} = \tilde{v}_k^{(t)} + s_k.
    \label{eq:dpc_tx}
\end{equation}
Due to the dither, $\tilde{v}_k^{(t)}$ is uniformly distributed over $\mathcal{V}(\Lambda_c)$ and independent of $s_k$.
Each user therefore transmits with average power $\Delta^2/6 + E_s$, which we also adopt as the power budget $P$ for all schemes.
\vspace{-1em}

\subsection{Receiver Design}

At the \ac{BS}, the data symbols are decoded slot by slot, starting from the \ac{LMMSE} estimate
\begin{equation}
    \hat{\bm{x}}^{(t)} = \Big(\bm{H}^\mathsf{H}\bm{H} + \frac{\sigma_w^2}{P}\mathbf{I}_K\Big)^{-1}\bm{H}^\mathsf{H}\bm{y}^{(t)}.
    \label{eq:lmmse}
\end{equation}
The dithers are then added back, and the modulo operation of \eqref{eq:modL} undoes the precoding of \eqref{eq:dpc_precoding}, removing the computing symbols without the \ac{RX} having knowledge of them.
Quantizing the result to the fine lattice finally yields
\begin{equation}
    \hat{\bm{v}}^{(t)} = \mathrm{modL}\Big(Q_{\Lambda_f}\big(\mathrm{modL}(\hat{\bm{x}}^{(t)} + \bm{u}^{(t)},\, \Delta)\big),\, \Delta\Big),
    \label{eq:dpc_detection}
\end{equation}
where $\bm{u}^{(t)} \triangleq [u_1^{(t)}, \dots, u_K^{(t)}]^\mathsf{T}$, $Q_{\Lambda_f}(\cdot)$ denotes the nearest-neighbor quantizer of $\Lambda_f$, and $\mathrm{modL}(\cdot, \cdot)$ is applied element-wise.

Since $\bm{s}$ is constant over the block, it is estimated jointly from all $T$ slots, given the decoded data symbols.
To that end, the \ac{RX} re-applies the precoding of \eqref{eq:dpc_precoding} to $\hat{\bm{v}}^{(t)}$ for every candidate $\bm{s}$, yielding the candidate transmit signal
\begin{equation}
    \bar{\bm{x}}^{(t)}(\bm{s}) \triangleq \mathrm{modL}\big(\hat{\bm{v}}^{(t)} - \bm{s} - \bm{u}^{(t)},\, \Delta\big) + \bm{s},
    \label{eq:dpc_candidate}
\end{equation}
and selects the candidate that best matches the received signals,
\begin{equation}
    \hat{\bm{s}} = \arg\min_{\bm{s} \in \mathcal{S}^K} \sum_{t=1}^{T} \big\|\bm{y}^{(t)} - \bm{H}\bar{\bm{x}}^{(t)}(\bm{s})\big\|_2^2.
    \label{eq:dpc_ml}
\end{equation}
The target function is finally computed as $\hat{f} = \sum_{k=1}^{K} \hat{s}_k$.

\section{Learning to Compute on Dirty Paper}
\label{sec:deep_dpc}

While Lattice \ac{ICC} renders the data detection largely free of computing interference, it suffers from two limitations in computation.
First, the \ac{ML} search in \eqref{eq:dpc_ml} spans all $D^K$ candidates, with a complexity growing exponentially in $K$.
Second, since $s_k$ is subtracted before the modulo operation and added back after it, $s_k$ affects $x_k^{(t)}$ only through the multiple of $\Delta$ by which the modulo operation shifts its argument.
Computing symbols with the same shifts over all $T$ slots thus produce identical transmit signals for given data symbols and dithers, which are indistinguishable even without noise and cause an error floor of the \ac{MSE}.
Both limitations stem from the complete pre-cancellation, which obscures the signal of interest for computation.
We therefore propose Learned \ac{ICC}, in which the precoder and the receiver are replaced by a neural encoder and decoder, trained end-to-end to balance both tasks, as illustrated in Fig.~\ref{fig:architecture}.

\begin{figure*}[t]
\centering
\begin{tikzpicture}[
    font=\footnotesize, >=Stealth, thick,
    node_io/.style={align=center},
    block_nn/.style={rectangle, draw, rounded corners=1mm, fill=blue!10, text width=1.8cm, align=center, minimum height=1.1cm},
    block_lmmse/.style={rectangle, draw, fill=purple!15, text width=1.6cm, align=center, minimum height=1.1cm},
    block_head/.style={rectangle, draw, rounded corners=1mm, fill=green!15, text width=2.0cm, align=center, minimum height=0.9cm},
    sum/.style={circle, draw, fill=gray!10, inner sep=0pt, minimum size=5mm}
]
\node[node_io] (tx_in) {$\begin{bmatrix} v_k^{(t)} \\ s_k \\ u_k^{(t)} \end{bmatrix}$};
\node[block_nn, right=0.5cm of tx_in] (enc) {Encoder\\$f_{\mathrm{enc}}$};
\node[sum, right=0.6cm of enc] (tx_add) {$+$};
\node[node_io, below=0.5cm of tx_add] (sk_inject) {$s_k$};
\draw[->] (tx_in) -- (enc);
\draw[->] (enc) -- node[above] {$\tilde{v}_k^{(t)}$} (tx_add);
\draw[->] (sk_inject) -- (tx_add);
\begin{scope}[on background layer]
    \node[draw=gray, dashed, fill=gray!5, inner sep=2.5mm, fit=(tx_in) (enc) (tx_add) (sk_inject),
          label={[anchor=south, font=\footnotesize\bfseries]above:User $k$ (shared encoder)}] {};
\end{scope}
\node[sum, right=1.0cm of tx_add] (mac) {$+$};
\node[above=0.5cm of mac, node_io] (noise) {$\bm{w}^{(t)}$};
\node[below=0.5cm of mac, node_io] (interf) {Other $K-1$\\users};
\draw[->] (tx_add) -- node[above] {$\;\;x_k^{(t)}$} node[below] {$\;\;\bm{h}_k$} (mac);
\draw[->] (noise) -- (mac);
\draw[->] (interf) -- (mac);
\node[block_lmmse, right=1.4cm of mac] (lmmse) {Spatial hint\\(LMMSE)};
\draw[->] (mac) -- node[above, pos=0.2] {$\bm{y}^{(t)}$} coordinate[pos=0.55] (ybranch) (lmmse);
\node[block_nn, right=1.0cm of lmmse, text width=2.0cm, minimum height=2.6cm] (trunk) {Shared trunk\\$f_{\mathrm{shared}}$};
\draw[->] (lmmse) -- node[above] {$\hat{\bm{x}}^{(t)}$} (trunk);
\draw[->] (ybranch) |- ([yshift=0.9cm]trunk.west);
\node[node_io, below=0.7cm of lmmse] (context) {$\bm{H}, \bm{u}^{(t)}$};
\draw[->] (context.east) -- ++(0.5,0) |- ([yshift=-0.9cm]trunk.west);
\coordinate[right=0.7cm of trunk.east] (split);
\draw[thick] (trunk.east) -- node[above] {$\bm{c}^{(t)}$} (split);
\node[block_head, right=0.5cm of split, yshift=0.75cm] (head_data) {Data head\\(classification)};
\node[block_head, right=0.5cm of split, yshift=-0.75cm] (head_comp) {AirComp head\\(regression)};
\draw[->] (split) |- (head_data.west);
\draw[->] (split) |- (head_comp.west);
\fill (split) circle (1.2pt);
\node[node_io, right=0.4cm of head_data] (out_v) {$\hat{\bm{v}}^{(t)}$};
\node[node_io, right=0.4cm of head_comp] (out_f) {$\hat{f}^{(t)}$};
\draw[->] (head_data) -- (out_v);
\draw[->] (head_comp) -- (out_f);
\begin{scope}[on background layer]
    \node[draw=gray, dashed, fill=gray!5, inner sep=2.5mm, fit=(lmmse) (context) (trunk) (out_v) (out_f),
          label={[anchor=south, font=\footnotesize\bfseries]above:Base station (decoder)}] {};
\end{scope}
\end{tikzpicture}
\caption{Proposed Learned \ac{ICC} architecture. Each user applies the same neural encoder to its own data symbol, computing symbol and dither, while the \ac{BS} jointly decodes the data of all users and estimates the target function via a dual-headed neural decoder.}
\label{fig:architecture}
\vspace{-1em}
\end{figure*}

\subsection{Neural Network Model}

Both the encoder and the decoder are \acp{MLP}, consisting of $L$ fully connected layers, with $L$ and the layer widths specific to each network.
Given an input vector $\bm{a}_0$, specified for the encoder and the decoder in \eqref{eq:enc} and \eqref{eq:dec_input}, respectively, each hidden layer computes
\begin{equation}
    \bm{a}_\ell = \mathrm{GeLU}\big(\bm{W}_\ell\,\bm{a}_{\ell-1} + \bm{b}_\ell\big), \quad \ell = 1, \dots, L-1,
    \label{eq:mlp}
\end{equation}
where $\bm{W}_\ell$ and $\bm{b}_\ell$ denote the trainable weights and biases of the $\ell$-th layer, and $\mathrm{GeLU}(\cdot)$ is the \ac{GeLU} activation \cite{Hendryks_2023}, applied element-wise.
The output layer computes $\bm{a}_L = \bm{W}_L\,\bm{a}_{L-1} + \bm{b}_L$ without activation, such that the outputs can take arbitrary real values.
Each hidden layer thus applies an affine transformation followed by a nonlinearity, which allows the network to approximate nonlinear mappings such as the modulo operation of \eqref{eq:dpc_precoding}.
All complex-valued quantities are represented by their stacked real and imaginary parts.

\subsection{Encoder}

All users employ the same encoder $f_{\mathrm{enc}}(\cdot\,;\Theta_{\mathrm{enc}})$, an \ac{MLP} as in \eqref{eq:mlp}, whose parameters $\Theta_{\mathrm{enc}} \triangleq \{\bm{W}_\ell, \bm{b}_\ell\}_{\ell=1}^{L}$ are learned during training, so that a single model serves all users, independently of $K$.
Replacing the modulo operation of \eqref{eq:dpc_precoding}, the encoder computes the precoded data symbol from local inputs without \ac{CSI} or the symbols of other users as
\begin{equation}
    \tilde{v}_k^{(t)} = f_{\mathrm{enc}}\big([v_k^{(t)}, s_k, u_k^{(t)}]^\mathsf{T};\, \Theta_{\mathrm{enc}}\big),
    \label{eq:enc}
\end{equation}
where the dither is included for comparability with Lattice \ac{ICC}.
As in Lattice \ac{ICC}, the computing symbol is then superimposed, yielding $x_k^{(t)} = \tilde{v}_k^{(t)} + s_k$, such that the encoder determines, through $\tilde{v}_k^{(t)}$, how the data symbols should be written around the computing symbol.

\subsection{Decoder}

At the \ac{BS}, the received signal is first equalized as in \eqref{eq:lmmse}, and the resulting estimate $\hat{\bm{x}}^{(t)}$ is provided to the decoder as a spatial hint.
The hint relieves the decoder of learning a matrix inversion, leaving its capacity for the nonlinear task of undoing the precoding.
Since the \ac{LMMSE} filter depends only on $\bm{H}$, $\sigma_w^2$ and $P$, it acts as a fixed linear map through which gradients propagate during training.

The decoder input stacks the received signal, the vectorized channel, the spatial hint and the dithers into
\begin{equation}
    \bm{I}_{\mathrm{in}}^{(t)} \triangleq \big[\bm{y}^{(t)},\, \bm{H},\, \hat{\bm{x}}^{(t)},\, \bm{u}^{(t)}\big] \in \mathbb{R}^{2N + 2NK + 4K}.
    \label{eq:dec_input}
\end{equation}
A shared trunk $f_{\mathrm{shared}}(\cdot\,;\Theta_{\mathrm{dec}})$, also an \ac{MLP} as in \eqref{eq:mlp}, extracts the common features $\bm{c}^{(t)} \triangleq f_{\mathrm{shared}}\big(\bm{I}_{\mathrm{in}}^{(t)};\, \Theta_{\mathrm{dec}}\big)$, which are passed to two task-specific heads, with parameters also collected in $\Theta_{\mathrm{dec}}$.

The data head treats data detection as classification, with a linear layer mapping $\bm{c}^{(t)}$ to $KD$ logits $\bm{o}^{(t)} \in \mathbb{R}^{KD}$, which are converted into symbol probabilities via a per-user softmax,
\begin{equation}
    \hat{\bm{p}}_k^{(t)} = \mathrm{softmax}\Big(\bm{o}^{(t)}_{(k-1)D+1\,:\,kD}\Big) \in \mathbb{R}^{D},
    \label{eq:softmax}
\end{equation}
where $[\hat{\bm{p}}_k^{(t)}]_m$ is the predicted probability that user $k$ transmitted the $m$-th symbol of $\mathcal{D}$.
The most probable symbol is selected as the decision $\hat{v}_k^{(t)}$.

The \ac{AirComp} head treats the computation as regression, outputting the real and imaginary part of the per-slot estimate $\hat{f}^{(t)}$ of $f(\bm{s})$, without recovering the individual computing symbols, unlike the \ac{ML} estimator in \eqref{eq:dpc_ml}.
At inference, the $T$ per-slot estimates are averaged and mapped to the nearest valid sum,
\vspace{-1em}
\begin{equation}
    \hat{f} = \arg\min_{f \in \mathcal{F}} \bigg|\, f - \frac{1}{T}\sum_{t=1}^{T}\hat{f}^{(t)} \bigg|,
    \label{eq:snap}
\end{equation}
where $\mathcal{F} \triangleq \big\{\sum_{k=1}^{K} s_k : s_k \in \mathcal{S}\big\}$ denotes the set of all possible sums.

\subsection{Loss Function}
\label{sec:training}

The encoder and decoder are trained end-to-end on the weighted loss
\begin{equation}
    \mathcal{L}_{\mathrm{total}} = \lambda_1\mathcal{L}_{\mathrm{data}} + \lambda_2\mathcal{L}_{\mathrm{sum}} + \lambda_3\mathcal{L}_{\mathrm{power}},
    \label{eq:loss_total}
\end{equation}
where $\lambda_1$, $\lambda_2$ and $\lambda_3$ balance data detection, computation and transmit power, with each term averaged over a mini-batch.
The data loss is the \ac{CE}
\begin{equation}
    \mathcal{L}_{\mathrm{data}} = -\frac{1}{K}\sum_{k=1}^{K}\log\,[\hat{\bm{p}}_k^{(t)}]_{m_k}, 
    \label{eq:loss_data}
\end{equation}
a differentiable surrogate for the \ac{SER}, where $m_k$ is the index in $\mathcal{D}$ of the symbol transmitted by user $k$.
The computing loss is the \ac{MSE} of the sum estimate, normalized by the computing power,
\begin{equation}
    \mathcal{L}_{\mathrm{sum}} = \frac{1}{K E_s}\,\bigg|\,\hat{f}^{(t)} - \sum_{k=1}^{K} s_k\bigg|^2.
    \label{eq:loss_sum}
\end{equation}
Since no modulo operation bounds the transmit signal, the penalty
\begin{equation}
    \mathcal{L}_{\mathrm{power}} = \Big(\mathbb{E}\big[|x_k^{(t)}|^2\big] - P\Big)^2,
    \label{eq:loss_power}
\end{equation}
with the expectation taken over all users, acts as a soft power constraint.
As the channel in \eqref{eq:rx_signal} is linear in $\bm{x}^{(t)}$, the gradients of \eqref{eq:loss_total} propagate through the channel into the encoder, which thus learns its precoding from the errors observed at the \ac{BS}.

\section{Numerical Results}
\label{sec:results}

We consider $K = 2$ users and a \ac{BS} with $N = 5$ antennas, with blocks of $T = 5$ slots.
Data symbols are drawn from a \ac{QPSK} constellation, with $D = 4$, and computing symbols from its $45^\circ$-rotated version, with $E_d = E_s = 0.5$.
The coarse lattice parameter is set to $\Delta = 2$, for which $\mathcal{D}$ has average power $E_d$, and the power budget is $P = \Delta^2/6 + E_s = 7/6$ per user, with the noise variance set as $\sigma_w^2 = P/\mathrm{SNR}$ for all schemes.
The encoder of Learned \ac{ICC} has three hidden layers of $64$, $64$ and $32$ neurons, while the shared trunk of the decoder has three hidden layers of $256$, $256$ and $128$ neurons, followed by a linear data head and an AirComp head with one hidden layer of $64$ neurons.
Training is performed with the Adam optimizer \cite{Kingma2015adam} with learning rate $10^{-3}$ for $5{,}000$ iterations, each on a mini-batch of $1{,}024$ freshly generated samples of channels, data symbols, computing symbols, dithers and noise, at a single \ac{SNR} of $10$\,dB, with $\lambda_1 = \lambda_2 = 2$ and $\lambda_3 = 5$.
All schemes are evaluated on independently generated test data, averaged over $10^5$ Monte Carlo realizations per \ac{SNR} point.

We compare Learned \ac{ICC}, evaluated on a single slot ($T = 1$) and with averaging over the block ($T = 5$), with Lattice \ac{ICC} \cite{Ranasinghe_DPC_2026}, and with the \ac{SotA} \cite{Ranasinghe_2026}, termed Superposition \ac{ICC}.
The latter superimposes the data and computing symbols as in \eqref{eq:sota_tx} at power $P$, detects the data via \ac{LMMSE} while treating the computing signal as noise, and estimates the sum from the residual, averaged over the block and mapped to the nearest valid sum as in \eqref{eq:snap}.
As a reference for data detection, we include \ac{QPSK} with $E_d = 0.5$ without any computing signal.
Performance is measured by the \ac{SER} and the $\text{MSE} =\mathbb{E}\big[|\hat{f} - f(\bm{s})|^2\big]$.

\begin{figure}[t!]
\centering
\includegraphics[width=\linewidth]{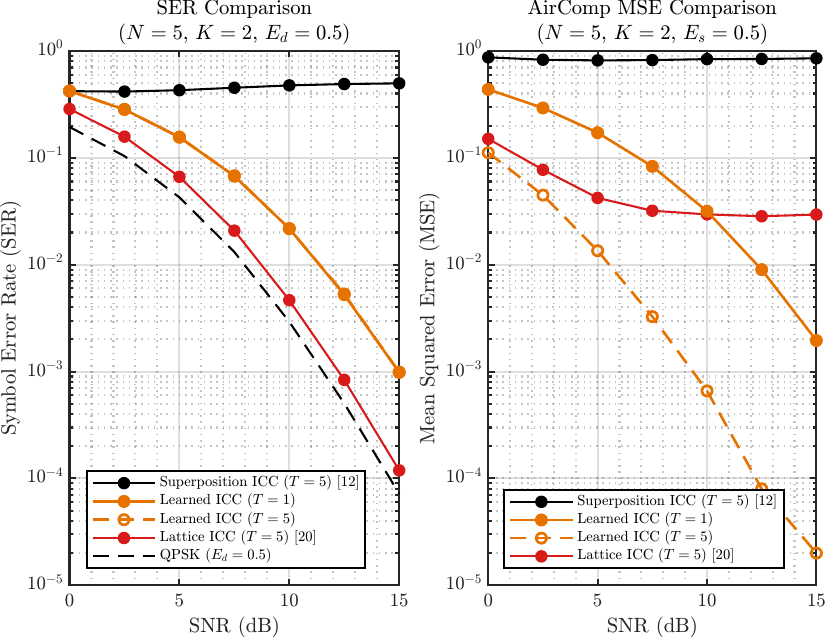}
\caption{SER (left) and MSE (right) of all schemes, for $N = 5$, $K = 2$ and $T = 5$.}
\vspace{-2em}
\label{fig:results}
\end{figure}

Fig.~\ref{fig:results} compares the \ac{SER} and the \ac{MSE} of all schemes.
Superposition \ac{ICC} is interference-limited in both tasks, since the computing symbols, treated as noise, are too large relative to the data symbols, and therefore, neither its \ac{SER} nor its \ac{MSE} improves with the \ac{SNR}.

Lattice \ac{ICC} removes the computing interference from the data, and its \ac{SER} closely follows the interference-free reference.
Its \ac{MSE}, however, saturates at moderate \acp{SNR}, since computing symbols that yield the same modulo shifts over the block are indistinguishable at the \ac{RX}.
At low \ac{SNR}, its \ac{ML} estimator additionally suffers from detection errors, as it relies on the decoded data.
Learned \ac{ICC}, whose encoder transmits with an average power of approximately $1.19$, close to the budget $P$, overcomes both effects when averaging over the block, outperforming Lattice \ac{ICC} across the entire \ac{SNR} range without an error floor.
Even on a single slot, it surpasses Lattice \ac{ICC} at high \ac{SNR}.
We attribute this gain to the learned precoding, which is not constrained to cancel the computing symbols entirely, and to the joint estimation of the data and the target function at the decoder.
In return, the uncancelled computing symbols share the complex plane with the data, reducing the distance between data symbols and causing a moderate \ac{SER} loss, identical for both $T$, compared to Lattice \ac{ICC}, while remaining far below Superposition \ac{ICC}.
In our experiments, the operating point on this trade-off is set by the loss weights in \eqref{eq:loss_total}, which determine how the encoder arranges the data and computing symbols in this plane.
Finally, although trained at a single \ac{SNR}, Learned \ac{ICC} performs well across the entire evaluated \ac{SNR} range, and, unlike the \ac{ML} search of Lattice \ac{ICC}, the complexity of its decoder does not grow exponentially with $K$.
\vspace{-1em}

\section{Conclusion}
\label{sec:conclusion}

We proposed Learned \ac{ICC}, a learning-based approach to \ac{ICC} on dirty paper, in which a neural encoder shared by all users replaces the lattice-based precoder, and a dual-headed neural decoder jointly detects the data and estimates the target function.
Simulation results showed that Learned \ac{ICC} removes the error floor of Lattice \ac{ICC} and substantially reduces the computation error, at a moderate cost in data detection.
Extending the approach to larger systems is left for future work.
\vspace{-1em}

\bibliographystyle{IEEEtran}
\bibliography{references}

\end{document}